# Direct observation of moiré flat-band breakdown at the edge of magic-angle twisted bilayer graphene

Long-Jing Yin[1,2]*, Ling-Hui Tong[1,2], Yue-Ying Zhou[1,2], Yang Zhang[1,2], Yuan Tian[1], Li Zhang[1], Lijie Zhang[1] and Zhihui Qin[1]*

[1] *Key Laboratory for Micro/Nano Optoelectronic Devices of Ministry of Education & Hunan Provincial Key Laboratory of Low-Dimensional Structural Physics and Devices, School of Physics and Electronics, Hunan University, Changsha 410082, China*

[2] *Research Institute of Hunan University in Chongqing, Chongqing 401120, China*

*Corresponding author: yinlj@hnu.edu.cn; zhqin@hnu.edu.cn

**Low-energy moiré flat bands in magic-angle twisted bilayer graphene (tBG) have demonstrated incredible potentials to exhibit rich exotic quantum phenomena. Theoretically, the moiré flat bands of tBG are based on the extended structures, i.e., the moiré patterns with periodic boundary conditions. However, a fundamental question of whether the flat bands can exist in the graphene moiré patterns with a reduced structure symmetry, such as sample edges, remains unanswered. Here, via scanning tunneling microscopy and spectroscopy, we study the local electronic properties of a magic-angle tBG near the sample terminated edge and report a direct observation of breakdown of the moiré flat bands. We show that the moiré electronic structures, including the low-energy flat bands, can sufficiently exist in a complete moiré spot, i.e., a moiré supercell, right at the edge even the translational symmetry of the moiré patterns is broken in one direction. However, the flat-band characteristic is obviously absent in the incomplete moiré spots that are partly terminated by the edge. Our results indicate that a whole moiré spot is sufficient and indispensable for the generation of the effective moiré flat bands in tBG.**

Magic-angle twisted bilayer graphene (tBG) has been extensively demonstrated to be a perfect platform for exploring a variety of exotic quantum phases, including strongly correlated and topological states [1-11]. For the emergence of these states, the existence of extremely flat electronic bands near the Fermi energy of the system plays a critical role [12]. In magic-angle tBG, such low-energy flat bands are created by the interlayer hybridization of the two component monolayers which are spatially modulated by the resulting moiré superlattices, and they are highly localized on the AA-stacking moiré spots regions. Therefore, in theory, the band structure of flat bands in tBG is determined by the extended moiré pattern, i.e., a superlattice with periodic boundary conditions [13-17]. Experimentally, the flat-band feature has been widely obtained in the periodic moiré patterns of tBG by measuring the resulting largely enhanced density of states (DOS) near the Fermi energy via scanning tunneling microscopy and spectroscopy (STM and STS) [18-24]. However, for a nonperiodic graphene moiré superlattice, or a moiré pattern with reduced structure symmetries, such as sample edges, whether the moiré flat bands can persist is still an open question.

In this work, we report STM and STS measurements near the sample terminated edge of a magic-angle tBG and present a direct observation of breakdown of the moiré flat bands. We find that the moiré electronic structures containing the low-energy flat bands can effectively exist in a complete moiré spot, i.e., a moiré supercell, right at the sample edge even the translational symmetry of the superlattices is broken in one direction. Remarkably, the flat-band feature—the sharp DOS peak—is strongly suppressed and even vanishes in the incomplete moiré spots that terminated by the edge, suggesting the breakdown of the flat bands. The results reported here demonstrate that to produce the effective moiré flat bands in tBG, a single, whole moiré spot is sufficient and needed at least, supporting a moiré quantum well picture.

Our experiments were performed on highly oriented pyrolytic graphite (HOPG, ZYA grade) substrate obtained from SurfaceNet GmbH. The HOPG samples were repeatedly surface cleaved by adhesive tape under air conditions and then quickly loaded into the STM chamber. HOPG naturally contains differently stacked domains including slightly twisted graphene layers. The topmost graphene layers with a twist, such as twisted bilayer and trilayer, are easily electronically decoupled from the HOPG bulk owing to the increased interlayer spacing between the top flake and underlying substrate after exfoliation, which has been previously reported in many experiments [18,25-28]. STM and STS measurements were carried out in a low-temperature (liquid nitrogen) STM system (CreaTec) under ultrahigh vacuum (~$10^{-11}$ Torr) and constant-current mode. The STM topographic images were calibrated against the standard graphene lattice, Si(111)-(7×7) lattice, and Au(111) surface. The STS (*dI/dV-V*) measurements were acquired by using a standard lock-in technique (AC bias modulation: 793 Hz and 10-20 mV). The electrochemically etched Pt/Ir (80%/20%) tips were used for the STM and STS measurements.

Figure 1(a) shows a representative STM topographic image near a step edge of the studied tBG. Clear moiré patterns are visualized in the image, where the bright spots are AA stacked sites and the dark regions are AB/BA stacked sites. The AB/BA stacked regions exhibit equal intensities in the STM image, suggesting the decoupling between the tBG and the underlying graphite substrate [29]. The wavelengths *L* of the moiré patterns measured in three superlattice directions are of 12.32 nm, 12.29 nm, and 12.65 nm, indicating negligible heterostrain in the sample [20]. We then obtain a twist angle $\theta$ of $1.13 \pm 0.02°$ determined from the relation between $\theta$ and *L*: $\theta = 2\arcsin(a/L)$, where $a = 0.142$ nm is the carbon–carbon distance. Such an angle is consistent well with the magic-angle in tBG, i.e., 1.1° [1,2,18]. A step edge with one layer height exists in the left region of Fig.1(a). The measured step height ~0.35 nm [obtained from the spacing between the lowest sites of the moiré patterns and the underlying layer, see the inset of Fig. 1(a)] is slightly larger than the equilibrium interlayer spacing (~0.33 nm) of the Bernal stacked graphene layers, suggesting the twist-induced weak coupling between the two layers of the tBG [26,27,30-32]. Below the step, the graphene region displays a clear honeycomb lattice structure [see the atomic-resolution STM image in Fig. 1(a)], which is a hallmark of monolayer graphene (see more data and discussion in Fig. S1 [33]). The above results demonstrate that the measured tBG sheet is effectively decoupled from the HOPG substrate (further evidence can be obtained from the STS measurements discussed below) [25].

Figure 1(b) shows a typical STS curve, i.e., the differential conductivity (*dI/dV*) spectrum which is proportional to the local density of states (LDOS), measured in the AA region of the tBG. The obtained *dI/dV* spectrum exhibits a strong and sharp DOS peak located close to the Fermi level. Flanking this sharp peak at ~±80 meV, two smaller side peaks grow. From the measured V-shaped STS spectrum of the monolayer graphene region below the step [blue curve in the inset of Fig. 1(b)], we can obtain that the Dirac point of the sample lies nearly at the Fermi level. Therefore, the sharp DOS peak at the Fermi energy is attributed to the moiré flat bands at the charge neutrality point of magic-angle tBG, and the two side peaks originate from the remote bands in the hole and electron sides respectively. To further reveal the electronic behaviors of the moiré bands, we study the spatial distributions of the electronic states by measuring site-resolved STS spectra over several moiré supercells. The result is shown in Figs. 1(c) and 1(d). Obviously, the electronic structures of the tBG are dramatically modulated by the moiré periodicity. Remarkably, the flat-band electronic states are highly localized on the AA stacking sites of the moiré patterns. However, the electronic states at high energies display opposite behaviors: it shows the lowest DOS in the AA regions. The above spectroscopic results—including the peak features and the spatial evolutions—are consistent well with the theoretical description of the moiré bands for magic-angle tBG [13,16] and also with previous STM reports [18-21], indicating that our sample indeed behaves as a magic-angle tBG and is electronically decoupled from the substrate. It's worth noting that there is no clear splitting of the flat-band peak in our sample. As previously reported in magic-angle tBG, the flat-band peak will split when it is located close to the Fermi energy (i.e., the case in our experiment) due to the enhanced electron-electron interactions that would result in correlated insulator states [19-21]. The reason why we do not see such a splitting in our tBG may be because of the suppression of the electron-electron interactions by the metallic graphite screening substrate [18,34], or due to the limitation of the spectroscopic energy resolution (~20 meV in our experiment from the equation $3k_BT$, where $k_B$ is the Boltzmann constant).

We now investigate the moiré band structures near the sample terminated edge of our 1.13° magic-angle tBG. Figure 2(a) shows the zoom-in STM topographic image of the step edge in Fig. 1(a). It displays that the step edge terminates the moiré periodicity with creation of some incomplete moiré bright spots at the edge. Such incomplete moiré spots provide an unprecedented platform for exploring the flat-band electronic properties in nonperiodic graphene moiré superlattice. Figure 2(b) shows the position-dependent STS spectra measured at three AA sites approaching the edge. The tunneling spectra [yellow and green curves in Fig. 2(b)] recorded in the two complete moiré AA spots near the edge exhibit almost the same spectroscopic features as that obtained at the AA sites away from the edge [Figs. 1(b) and 1(c)], i.e., a sharp flat-band peak at zero-energy accompanying with two small flanking peaks. However, the sharp flat-band DOS peak is strongly suppressed and hardly visualized in the STS spectrum [blue curve in Fig. 2(b)] taken at the incomplete moiré AA site that terminated by the edge, suggesting the destruction of the flat-band structures. To better reveal the evolution of the moiré flat bands near the edge, we measure the spatially resolved contour plot of *dI/dV* spectra close to the edge, as shown in Fig. 2(c) (see more data in Fig. S2 [33]). It demonstrates that although the high energy electronic states seem still follow the modulation of the moiré periodicity, the zero-energy flat-band peak is almost disappeared in the incomplete AA region at the edge. Such a phenomenon is reflected more obviously in the *dI/dV* profile line taken at the energy of the flat bands [as shown in Fig. 2(d)]: the DOS intensity of flat bands is greatly diminished at the incomplete AA site and even smaller than that at the saddle-point site. The above results indicate that the low-energy moiré flat bands break down in the incomplete moiré supercell.

To further explore the electronic structures of magic-angle tBG near the sample edge, we next turn to the LDOS dependence on terminated moiré supercells with various residual AA areas. Figure 3 shows the STS spectra measured on the moiré bright spots right at the edge with different terminated ratios. The sharp flat-band DOS peak preserves in the complete AA region right beside the edge [see the top curve of the ~1.0 spot (the number corresponds to the ratio of the measured moiré spot area to a whole one, see Fig. S3 for details [33]) in Fig. 3], indicating that the moiré flat bands can effectively exist in the intact moiré supercells at the sample edge even the translational symmetry of the superlattices is broken in one direction [35]. However, the low-energy moiré flat-band characteristic breaks down for those incomplete AA spots at the edge: the sharp zero-energy flat-band peak is absent in the STS spectra of ~0.9, ~0.7, and ~0.5 spots (see the lower three curves in Fig. 3). Markedly, the *dI/dV* spectrum taken at the AA site of the ~0.5 spot exhibits nearly a V-shaped feature, suggesting the thorough extinction of the low-energy flat bands. These AA-area-varied LDOS as well as the spatially resolved spectra described above thus all demonstrate the breakdown of the moiré flat-band structures near the sample edge. Note that the STS spectra of Fig. 3 are all measured at the positions out of the penetration range of the edge electronic states [36] (see Fig. S4), and we obtain distinct spectra at the same distance from the edge boundary for the complete and incomplete AA spots (see Fig. 3). Therefore, the edge effects—including the influences of electronic states and symmetry breaking at the edge—can be excluded in our experiment (see more discussion in Supplemental Material [33]), and the observed breakdown of the flat bands in incomplete AA spots can be attributed to the intrinsic property of the moiré superlattices.

Theoretically, the moiré flat bands of magic-angle tBG are produced by the extended structures, i.e., the moiré superlattices with periodic boundary conditions [13-17]. This result has been extensively verified in the periodic moiré patterns of tBG by transports [1,2] and spectroscopic experiments [18-24,37]. However, the situation of the flat bands in those nonperiodic or finite-size graphene moiré superlattices is still unknown. The local spectroscopic measurement presented here explicitly demonstrates that the moiré flat bands survive in the complete AA spots right at the sample edge even though the translational symmetry of the superlattices is absent in one direction, and more importantly they break down in the incomplete AA spots that cut by the edge. That is to say, when the periodicity within the lattice of a complete AA spot is preserved, the flat bands will exist in the supercell, even its translational symmetry is broken. This result implies that the moiré flat bands of tBG can be described by the physics of a single moiré supercell. Indeed, it has been predicted by the continuum model that each moiré supercell in tBG can be seen as a "moiré quantum well", in which the trapped electrons could produce the low-energy band structures analogous to that of the periodic supercells [35]. Our work gives the experimental support for such a moiré quantum well picture, and demonstrates that, to generate the effective low-energy flat bands in tBG, a single, whole moiré supercell is sufficient and also needed at least. In addition, the sample geometrical edge in a two-dimension material is highly related to many quantum phenomena of the system including edge states, edge transport, and quantum Hall effects [38-41]. The observation of the breakdown of moiré flat bands at the sample edge is thus instrumental in understanding of some fundamental issues, such as symmetry breaking and edge scattering, for those edge-modulated properties in magic-angle tBG.

In summary, the local electronic properties of the magic-angle tBG near the sample terminated edge were investigated by STM and STS measurements. We found that the moiré flat bands could survive in the complete AA spots right at the sample edge of tBG even the superlattice periodicity is disrupted in one direction. However, when the AA site is cut partly by the edge, the low-energy flat-band characteristic breaks down. These findings strongly imply that an intact moiré supercell is sufficient and also essential to create the effective zero-energy flat bands for tBG. Our work provides further fundamental knowledge about microscopic properties of graphene moiré flat-band structures and emphasizes the influence of edges on the electronic structures of twisted van der Waals systems.

## Acknowledgements

This work was supported by the National Natural Science Foundation of China (Grant Nos. 12174095, 11804089, 12174096, 51772087, 11904094 and 51972106), the Natural Science Foundation of Hunan Province, China (Grant No. 2021JJ20026), and the Strategic Priority Research Program of Chinese Academy of Sciences (Grant No. XDB30000000). L. J. Y. also acknowledges support from the Science and Technology Innovation Program of Hunan Province (Grant No. 2021RC3037) and the Natural Science Foundation of Chongqing, China (cstc2021jcyj-msxmX0381). The authors acknowledge the financial support from the Fundamental Research Funds for the Central Universities of China.

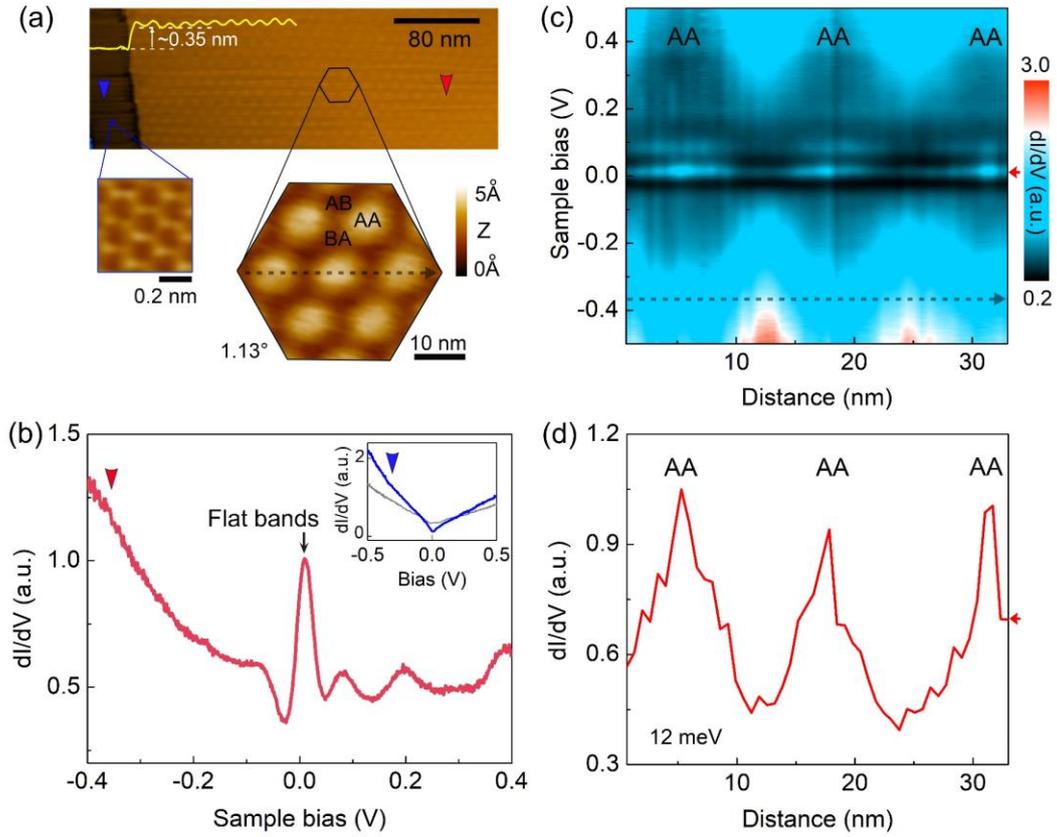

**FIG. 1.** (a) Top panel: STM topographic image ($V_b = 0.5$ V, $I = 0.2$ nA) of a 1.13° tBG near a step edge. Inset is the height profile of the step edge. Lower left panel: atomic-resolution STM image ($V_b = 0.3$ V, $I = 0.4$ nA) of the graphene layer below the step. Lower right panel: magnified STM image ($V_b = 42$ mV, $I = 0.2$ nA) obtained in the tBG region marked by the hexagon in the top panel. (b) Typical $dI/dV$-$V$ spectrum of the 1.13° tBG (AA site). Inset shows the STS spectra recorded in the graphene layer below the step of panel (a) (blue curve) and in the graphite substate (gray curve). It displays a finite DOS at the Dirac point for the spectrum of graphite, while a nearly zero DOS at the Dirac point for the spectrum of the monolayer region (a signature of massless Dirac Fermion), further indicating the decoupling feature of the studied sample. (c) Spatially resolved contour plot of $dI/dV$ spectra measured along the dashed arrow, i.e., the AA-(saddle-point of AB/BA)-AA direction, in the tBG region of panel (a). (d) $dI/dV$ profile line taken at the energy of 12 meV from panel (c).

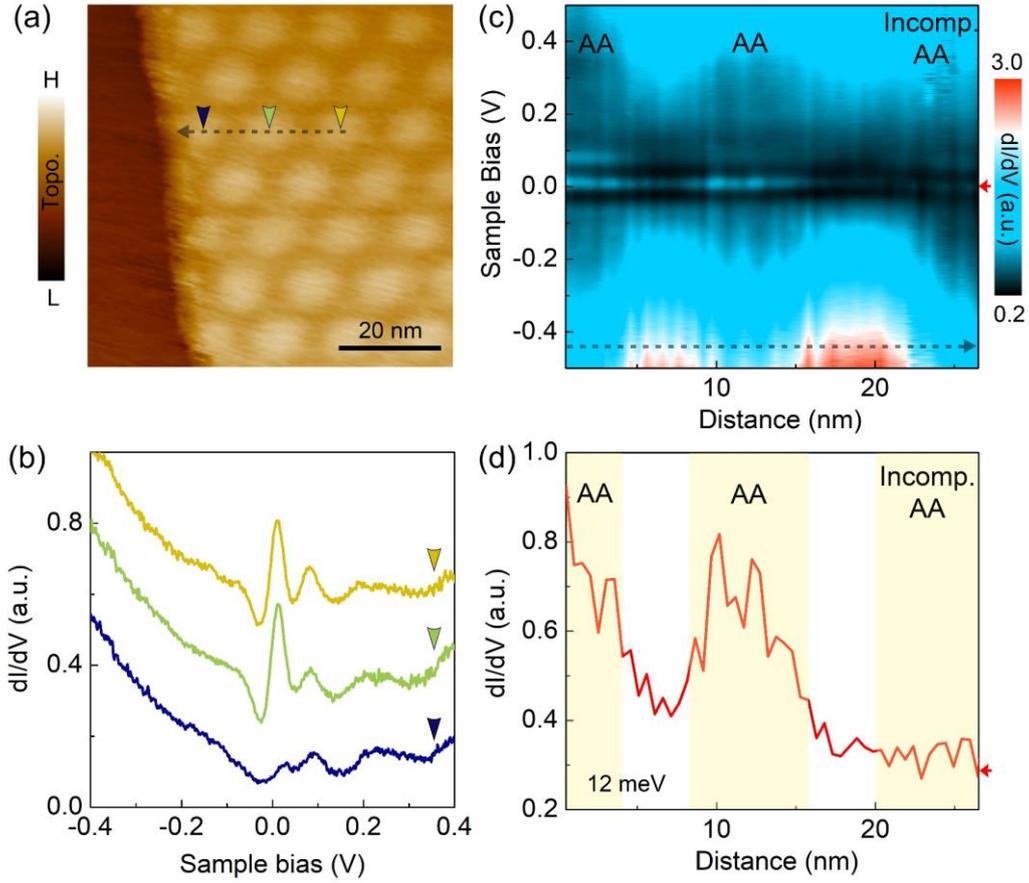

**FIG. 2.** (a) STM topographic image ($V_b$ = 0.3 V, $I$ = 0.1 nA) of the 1.13° tBG around the step edge. (b) STS spectra measured at the bright moiré spots (AA stacking regions) marked by the colored arrows in panel (a). The curves are offset in the *y* axis for clarity. (c) Spatially resolved *dI/dV* spectra obtained along the dashed arrow in the tBG of panel (a). (d) *dI/dV* profile line measured at the energy of 12 meV from panel (c).

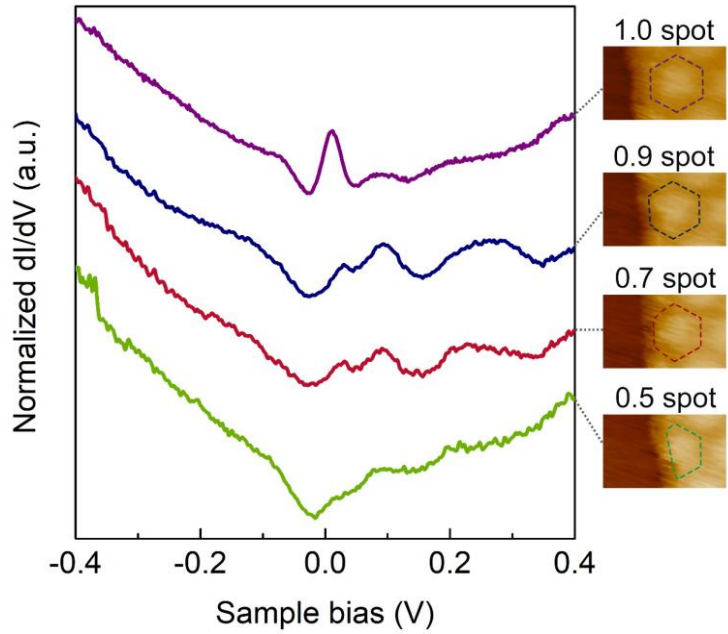

**FIG. 3.** Normalized *dI/dV* spectra recorded in the AA stacking regions, as marked by the polygons in the STM images of the right panels, with and without a complete moiré spot at the step edge. The measurement points are nearly at the same distance, i.e., ~4 nm for the 0.7-1.0 spots and ~3 nm for the 0.5 spot (within the AA region), from the edge boundary to avoid edge effects. The curves are offset in the *y* axis for clarity.